\documentclass[aps,prb,twocolumn,showpacs,superscriptaddress]{revtex4-1}
\usepackage{graphicx}
\usepackage{dcolumn}
\usepackage{bm}
\usepackage{amssymb}
\usepackage{amsmath}
\usepackage{subfigure}
\usepackage{float}

\begin{document}
\title{Monitoring Mechanical Motion of Carbon Nanotube based Nanomotor by Optical Absorption Spectrum}

\author{Baomin Wang}
\affiliation{School of Physics, Nankai University, Tianjin 300071, China}

\author{Xuewei Cao}
\email[]{xwcao@nankai.edu.cn}
\affiliation{School of Physics, Nankai University, Tianjin 300071, China}

\author{Zhan Wang}
\affiliation{School of Physics, Nankai University, Tianjin 300071, China}

\author{Yong Wang}
\email[]{yongwang@nankai.edu.cn}
\affiliation{School of Physics, Nankai University, Tianjin 300071, China}

\author{Kaihui Liu}
\affiliation{State Key Laboratory for Mesoscopic Physics, School of Physics, Peking University, Beijing 100871, China}

\begin{abstract}
The optical absorption spectrums of nanomotors made from double-wall carbon nanotubes have been calculated with the time-dependent density functional based tight binding method. When the outer short tube of the nanomotor moves along or rotates around the inner long tube, the peaks in the spectrum will gradually evolve and may shift periodically, the amplitude of which can be as large as hundreds of meV. We show that the features and behaviors of the optical absorption spectrum could be used to monitor the mechanical motions of the double-wall carbon nanotube based nanomotor.
\end{abstract}

\maketitle

\section{Introduction}
Manufacturing nanoelectromechanical systems (NEMS) is one of the central tasks of modern nanotechnologies, which promises potential advantages over classical devices with larger sizes.\cite{NEMS2002} Nanomotor is one typical and important NEMS which can convert other forms of energy into mechanical energy.\cite{Nanoscale2015} Several technique schemes have been exploited to fabricate nanomotors, and multi-wall carbon nanotubes are the natural and competitive candidates due to their special atomic structures.\cite{Nature2003,NanoLett2004,PRL200801,Science2008,ACSnano2010,Nanoscale2010,NanoLett2014} Considering a double-wall carbon nanotube (DWCNT) based nanomotor, the two ends of the inner long tube is fixed to two anchor pads, while the outer short tube can move along and/or rotate around the inner tube under the external driven electric or thermal field. This tiny nanomotor can be used as the key component to design nano devices for given purposes. For example, a cargo can be constructed by attaching a metal plate on the outer tube, which may be used to deliver drugs for medical applications.\cite{Science2008}

In the past studies, the mechanical motion of DWCNT based nanomotor is usually observed by the scanning electron microscope (SEM) apparatus.\cite{Nature2003,NanoLett2004,Science2008} Although the SEM images can clearly determine the various configurations of the nanomotor during the motion, it is hard for SEM apparatus to continuously capture the transient dynamics of the nanomotor, especially if the motion velocity reaches the prediction value $10^{8}$ $\mu$m/s as given by molecular dynamics simulation.\cite{Science2008} Therefore, alternative methods are required to probe the mechanical motion of the nanomotor. For example, Cai \emph{et al} have proposed to use the centrifugal effect to measure the high-speed rotation of a thermal carbon nanomotor.\cite{SciRep2016} Recently, Liu \textit{et al} have systematically developed the real-time optical imaging and broadband \emph{in situ} spectroscopy, and have studied the optical absorption spectrum of various types of CNTs.\cite{NatNano2013,NatPhys2014,PNAS2014} Their experimental results revealed that the interaction between two individual tubes can significantly modify the optical transition energy in DWCNTs.\cite{NatPhys2014} Their findings suggest that optical absorption spectrum can be helpful to identify the relative handedness of DWCNTs, which is even indistinguishable by the electron diffraction pattern.\cite{NatPhys2014} Therefore, it should be practical to monitor the mechanical motion of the nanomotor made from DWCNT by measuring the variation of the optical absorption spectrum.

In this paper, we study the optical absorption spectrum of the DWCNT based nanomotor with the time-dependent density functional based tight binding (TD-DFTB) method.\cite{PRB200101,THEOCHEM2009} The DFTB method takes certain approximate strategies to implement the density functional theory\cite{PRB1995,PRB1998} and has been widely applied to study different physics properties of carbon nanostructures such as carbon nanotube,\cite{PRL2000,JACS2001}fullerence,\cite{PRB200101,PRL200802}nanodiamond,\cite{Science2002,Nanoscale2011,Nanoscale2012,Nanoscale2014,Nanoscale2016} \textit{etc}. Below, we will show the evolution of the optical absorption spectrum of nanomotors when the outer tube moves along or rotates around the inner tube, which demonstrates the possibility to identify the given configuration of the nanomotor from its optical absorption spectrum and monitor its mechanical motion.
\begin{figure*}[ht!]
\centering
  \includegraphics[scale=0.7,clip]{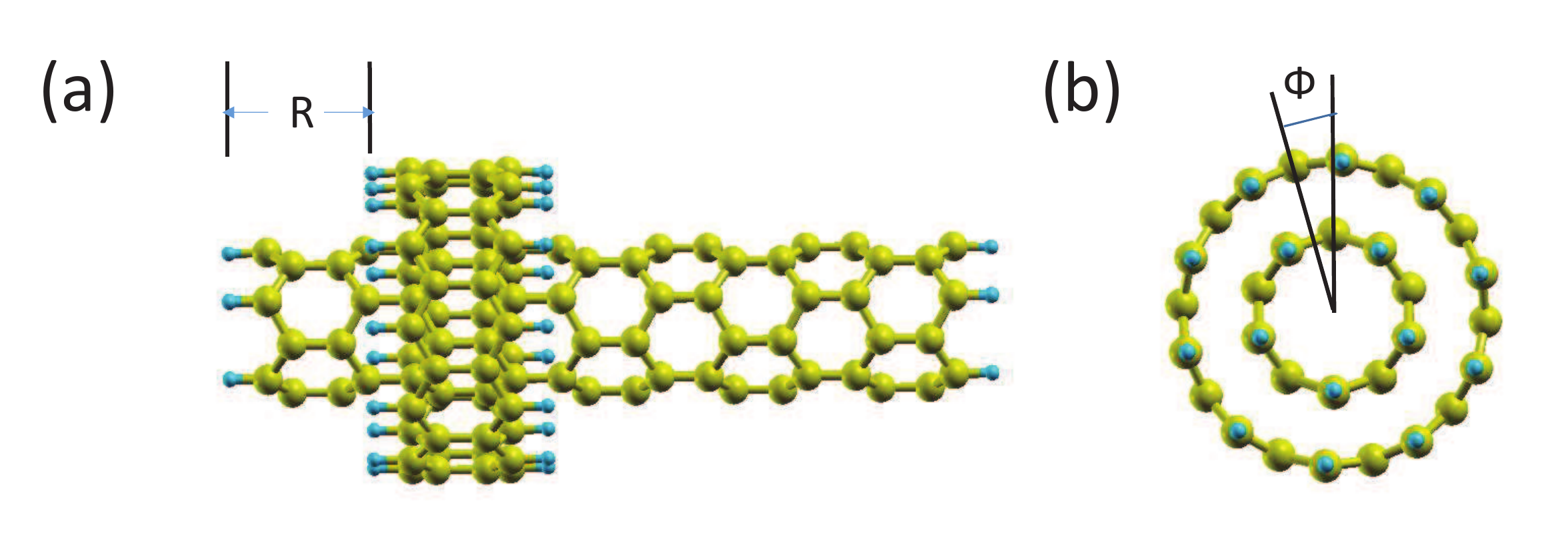}
  \caption{(Color online) Atomistic structure of a DWCNT based nanomotor. The ends of the carbon nanotubes are passivated by hydrogen atoms. (a) the translation distance $R$ of the outer tube moving along the axis of the inner tube; (b) the counterclockwise rotation angle $\Phi$ of the outer tube around the axis of the inner tube. Green : carbon atoms; blue : hydrogen atoms.}
  \label{struct}
\end{figure*}

\section{Structure and Method}
The typical structure of a DWCNT nanomotor is shown in Fig. 1, which consists of a long inner tube and a short outer tube. (5,0) CNT is chosen as the inner tube, and (10,0) or (11,0) CNT is chosen as the outer tube in our calculations. The ends of both the tubes are passivated by hydrogen atoms. The atomic positions of the two tubes are relaxed separately by the geometry optimization algorithm implemented in DFTB+ package\cite{JPCA2007} before they are assembled together. The deformation effect of the two tubes has been neglected due to the weakness of the Van der Waals interaction. For a given nanomotor, its configuration is characterized by the translation distance $R$ along the axis direction and the counterclockwise rotation angle $\Phi$ of the outer tube in relative to the inner tube, as defined in Fig.~\ref{struct}.

The optical absorption spectrum for each configuration is calculated with TD-DFTB method\cite{PRB200101,THEOCHEM2009}. In this approach, the ground state of the electronic structure is first obtained by performing the self-consistent DFTB calculation,\cite{PRB1995,PRB1998} which gives the corresponding Kohn-Sham (KS) orbitals $\psi_{i}$ and the KS energies $\epsilon_{i}$. Then the optical spectrum $\Omega_{I}$ is calculated from the Casida equation\cite{Casida1996}
\begin{equation}
\sum_{kl\tau}[\omega_{ij}^{2}\delta_{ik}\delta_{jl}\delta_{\delta\tau}+2\sqrt{\omega_{ij}}K_{ij\sigma,kl\tau}
\sqrt{\omega_{kl}}]F_{kl\tau}^{I}=\Omega_{I}^{2}F_{ij\sigma}^{I}\label{eigenequ}
\end{equation}
where the coupling matrix $K_{ij\sigma,kl\tau}$ is
\begin{eqnarray}
K_{ij\sigma,kl\tau}=\int\int&&\psi_{i}(\mathbf{r})\psi_{j}(\mathbf{r})(\frac{1}{|\mathbf{r}-\mathbf{r}'|}
+\frac{\delta^{2}E_{xc}}{\delta\rho_{\sigma}(\mathbf{r})\delta\rho_{\tau}(\mathbf{r}')})\nonumber\\
&\times&\psi_{k}(\mathbf{r}')\psi_{l}(\mathbf{r}')
d\mathbf{r}d\mathbf{r}'\label{Kmax}
\end{eqnarray}
and $\omega_{ik}=\epsilon_{j}-\epsilon_{i}$. $\sigma$ and $\tau$ denote the electron spin. Besides, the absorption strength for singlet-singlet transition is given by
\begin{equation}
f_{I}=\frac{2}{3}\Omega_{I}\sum_{k=x,y,z}|\sum_{ij\sigma}\langle\psi_{i}|\mathbf{r}_{k}|\psi_{j}\rangle\sqrt{\frac{\omega_{ij}}{\Omega_{I}}}F_{ij\sigma}^{I}|^{2}\label{OS}
\end{equation}
This method has been successfully applied to study the optical properties of various nanoscale systems including fullerence,\cite{PRB200101}silicon nanoclusters,\cite{APL2007,JPCC200701,JPCC2009,JPCC2014}cadmium sulfide\cite{JPCB2003,JPCC200702,JPCC2011} and zinc selenide nanostructures,\cite{PRB2006} \textit{etc}.

\section{Results and discussion}
\begin{figure}[ht!]
\centering
  \includegraphics[scale=0.5,clip]{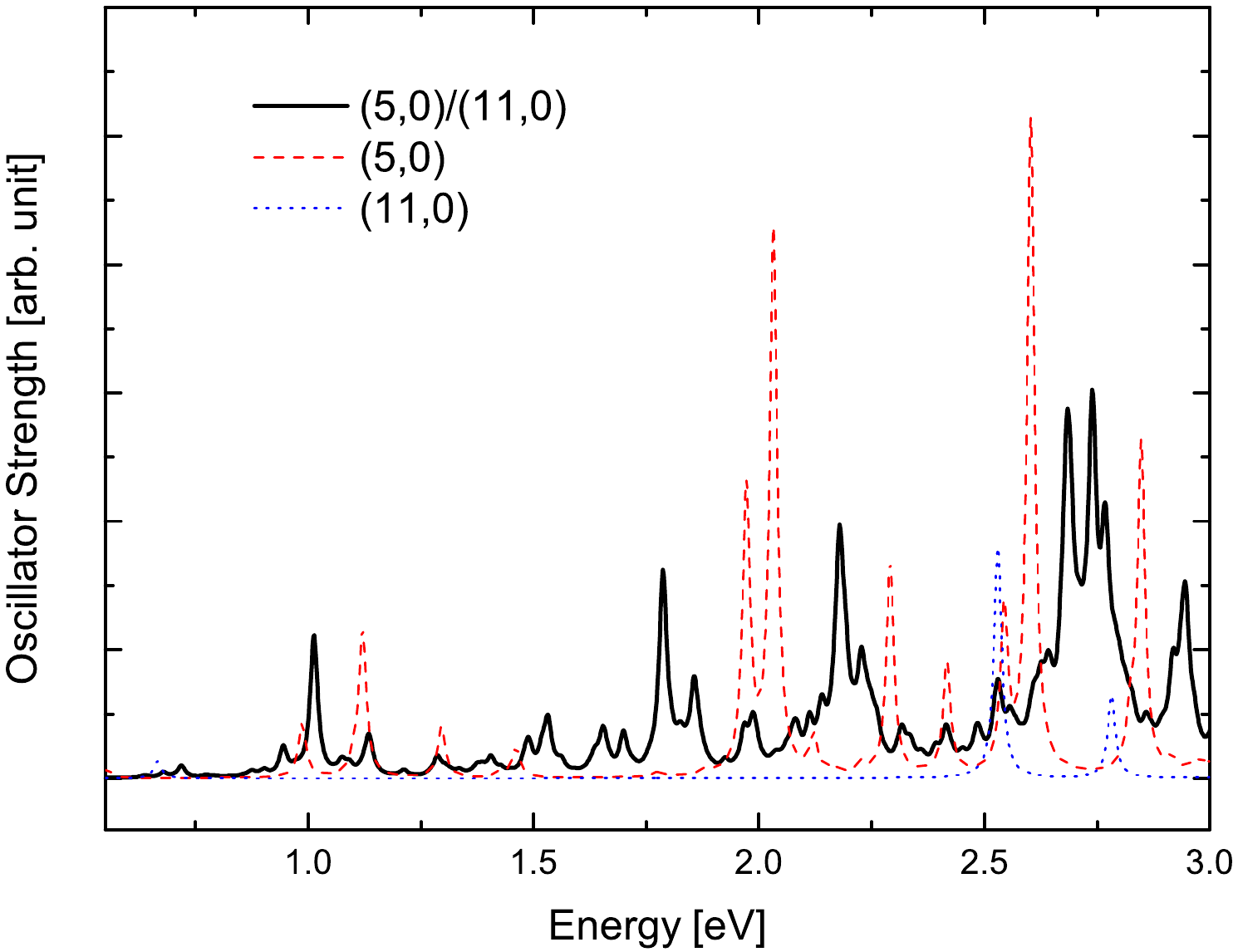}
  \caption{(Color online) The optical absorption spectrum of the nanomotor made from (5,0)/(11,0) DWCNT (black real line), and the spectrums for the constituent inner long tube (red dash line) and outer short tube (blue dot line). The configuration of the nanomotor is set as $R=0$ and $\Phi=0$. The spectrums are broaden by Lorentz lineshape with width $0.02$~eV.}
  \label{optsum}
\end{figure}
First, we investigate the effect of the coupling interaction between the two tubes in the nanomotor on the optical absorption spectrum. We consider a nanomotor made from the (5,0)/(11,0) DWCNT as an example, where the outer tube has one unit cell and the inner tube has five unit cells. Calculation for longer tubes is more realistic and is feasible by the TD-DFTB method in principle, but is limited by our computation power. The optical spectrum of ultrashort CNT will be seriously affected by the quantum confinement effect.\cite{JACS2008} The configuration of the nanomotor is set as $R=0$~\AA~and $\Phi=0^{\circ}$ in the calculation. The (5,0) CNT has been found to be metallic because of the $\sigma-\pi$ mixing induced by high curvature.\cite{PRB2002} Thus, a small gap about $47$~meV is obtained for the inner tube due to the quantum confinement effect. The results shown in Fig.~\ref{optsum} suggest that the optical absorption spectrum of the nanomotor is not a simple summation of the spectrums of the two constituent tubes, which implies strong inter-tube interaction in this structure. It is observed that much more peaks appear in the optical spectrum of the nanomotor in Fig.~\ref{optsum}. This can be understood from the lower symmetry of the nanomotor due to inter-tube interaction, which breaks the original selection rules for the quantum state transitions in individual tubes and makes more state transitions to be allowed. For comparison, we have also calculated the spectrum for another nanomotor made from (5,0)/(16,0) DWCNT with larger inter-tube separation and negligible inter-tube interaction, which is just the summation of the spectrums of the inner and outer tubes. Therefore, it is possible to infer the configuration of the nanomotor from its optical absorption spectrum only if there exists strong enough inter-tuber interaction.
\begin{figure}[ht!]
 \centering
 \includegraphics[scale=0.2,clip]{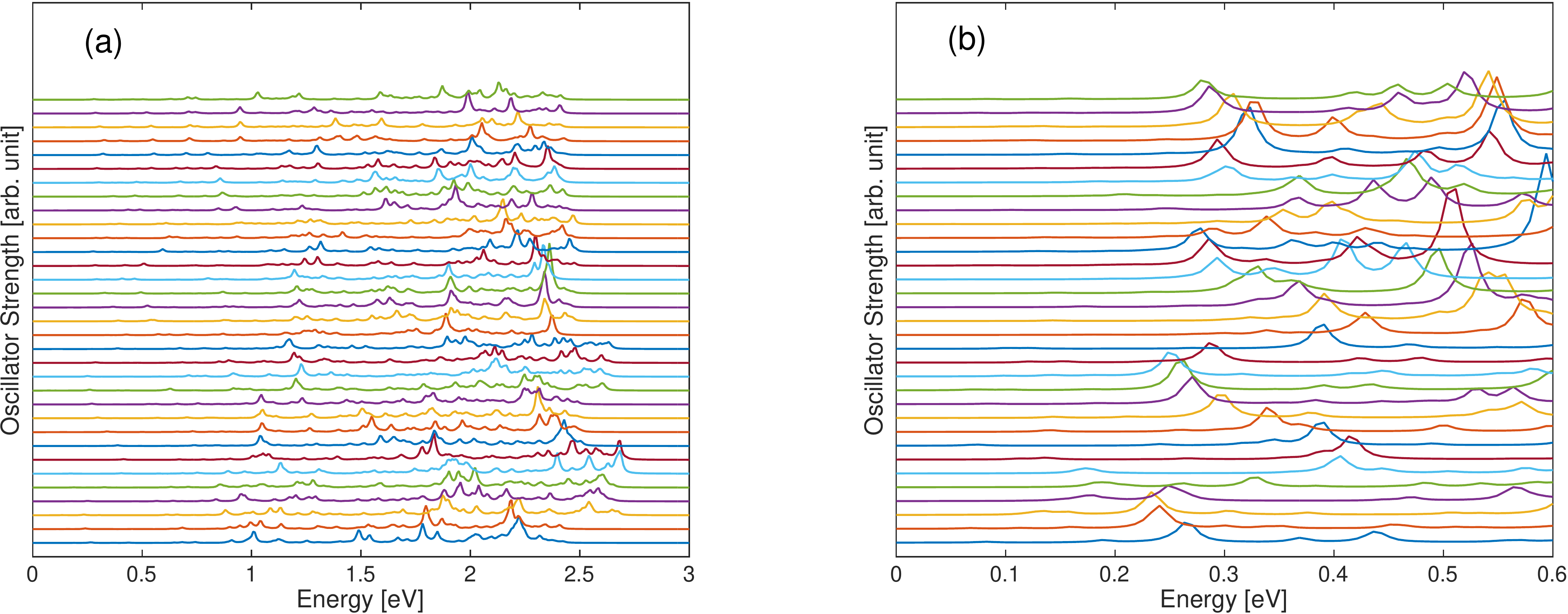}
 \includegraphics[scale=0.16,clip]{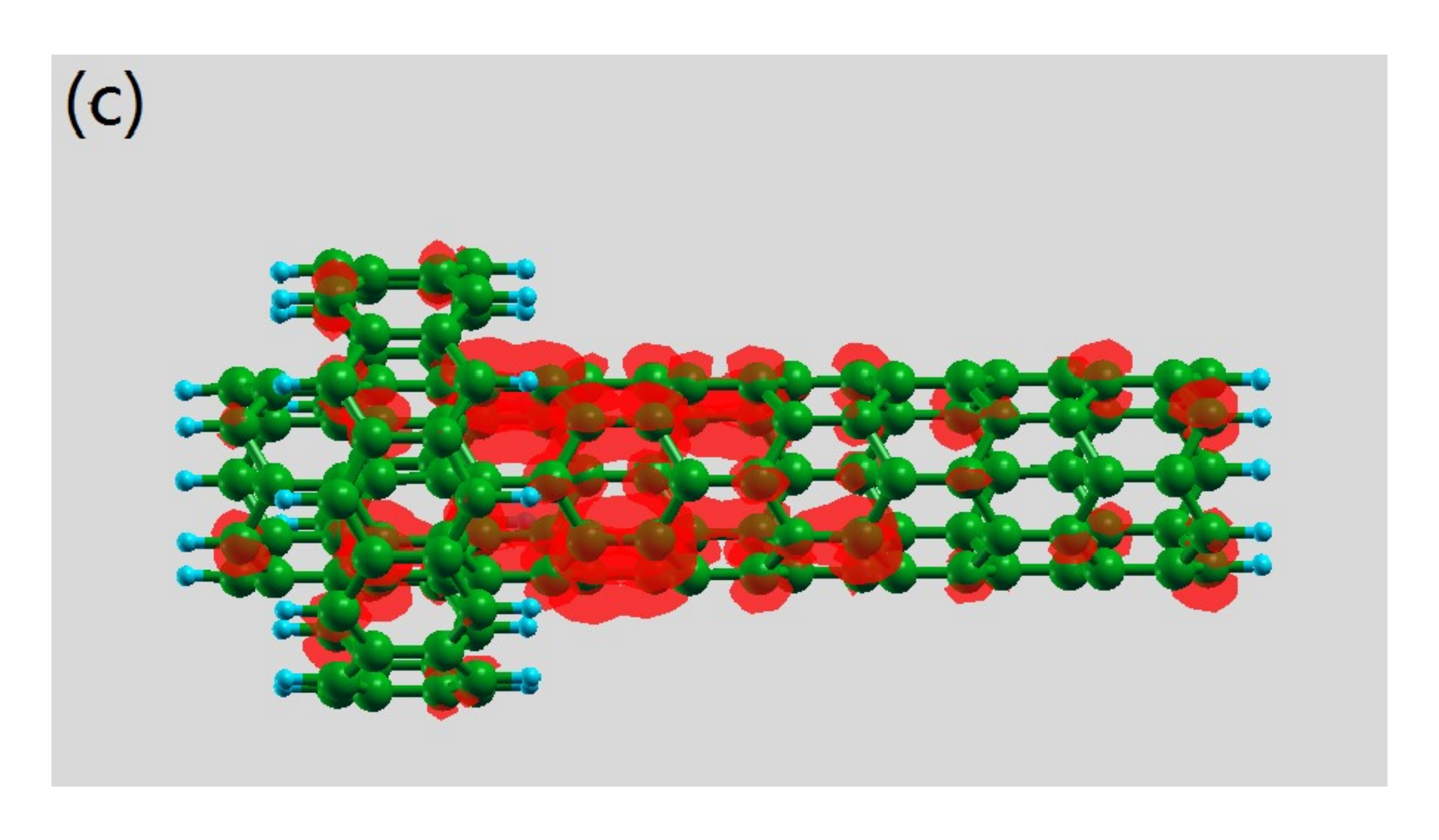}
 \includegraphics[scale=0.16,clip]{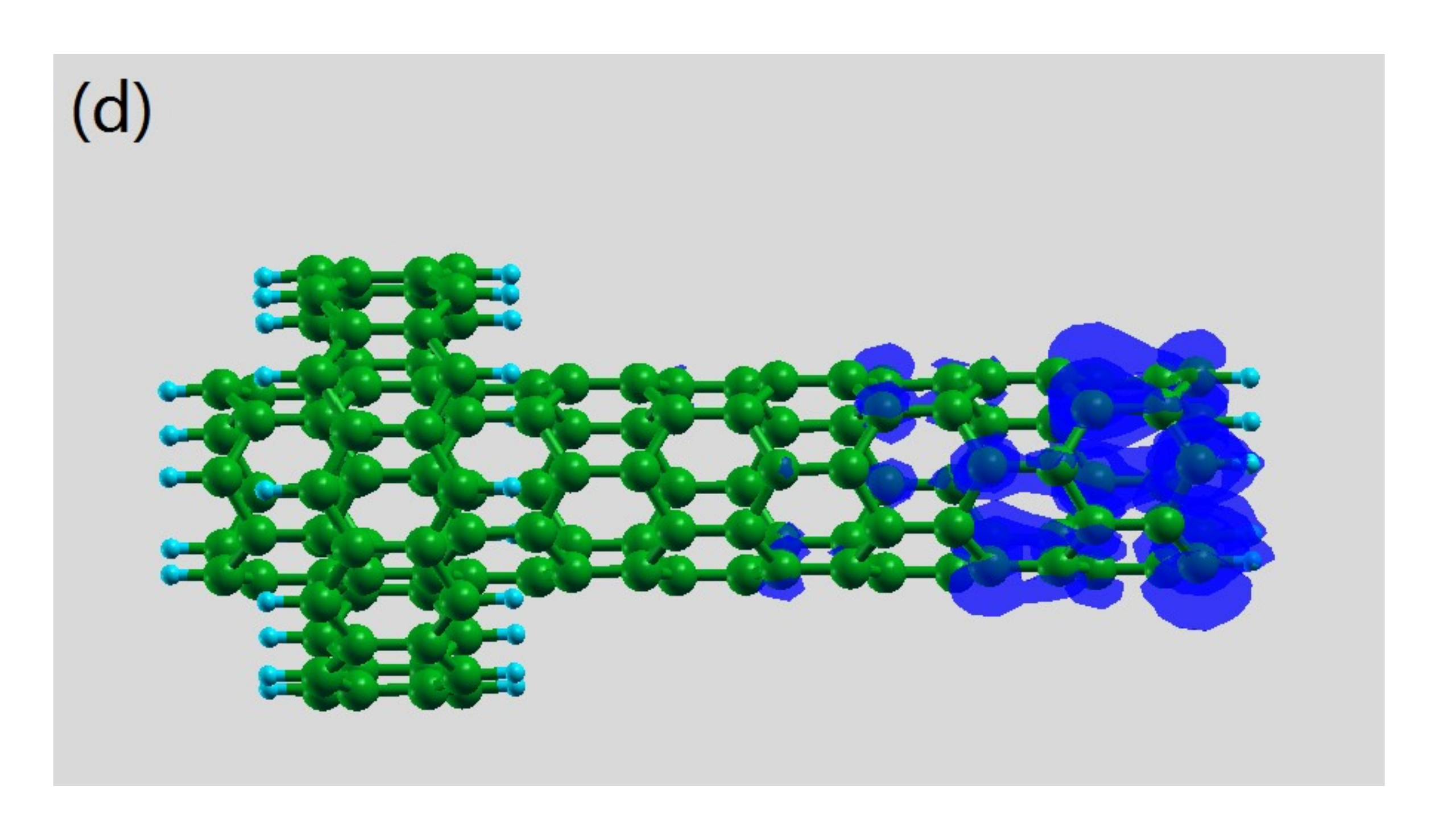}
 \caption{(Color online) (a)(b): the optical absorption spectrums of the nanomotor made from (5,0)/(11,0) DWCNT when the outer tube moves along the inner tube. Here, the rotation angle $\Phi$ is fixed as $0^{\circ}$, and the translation distance $R$ varies from $0$~\AA~to $8$~\AA~with step length $0.25$~\AA~(shown from bottom to top). The spectrums are broaden by Lorentz lineshape with width $0.02$~eV. (c)(d): charge density distributions of the initial (red) and final (blue) state of the transition corresponding to the first strong peak when $R=2$~\AA~in (b), and the isovalue is taken as $5\times10^{-4}$.}
 \label{transopt}
\end{figure}

We further calculate the optical absorption spectrum of the same nanomotor in Fig.~\ref{optsum} when translation distance $R$ is increased from $0$~\AA~ to $8$~\AA~with step length $0.25$~\AA~and the rotation angle $\Phi$ is fixed as $0^{\circ}$, and the results are shown in Fig.~\ref{transopt}(a) and (b). We find that the evolution of spectrums in Fig.~\ref{transopt}(a) lacks periodic pattern in terms of $R$, which is caused by the breaking of translation symmetry of the inner tube with finite length. However, after carefully examining the finer structure of the spectrums in the energy window [0,~0.6]~eV, we find that the first relatively strong peak shifts with the increasing $R$ in a periodic pattern approximately, as shown in Fig.~\ref{transopt}(b). The shift period is around $2$~\AA, which is about half of the lattice constant $a$ with $4.26$~\AA~for the (5,0) CNT here. This phenomenon can be understood from the fact that the crystal structure of the ideal (5,0) CNT can be recovered by the joint operations of $\frac{a}{2}$-translation and $\frac{\pi}{5}$-rotation. If the details of the atomic structure of the outer tube can be neglected, the $\frac{a}{2}$-period of the peak shift will approximately hold. This periodic shift is however destroyed for the optical absorption spectrum in higher energy range, which implies the higher sensitivity of the spectrum to the quantum confinement effect. Besides, we observe that the shift amplitude of the first strong peak is as large as $150$~meV, which can be easily detected by the modern optical spectroscopy technique.\cite{NatNano2013,NatPhys2014,PNAS2014} In Fig.~\ref{optsum} (c) and (d), we plot the charge density distributions of the initial and final states for the transition corresponding to the first strong peak when $R=2$~\AA~respectively. We find that the initial state is distributed in the middle of the inner tube; while the final state is located at the end of the inner tube and is far from the outer tube. Thus, the observed periodic shift of the peak here is caused by the modulation of the wavefunction of the initial state by the position of the outer tube, which in turn can be used to monitor the translational motion of the nanomotor.

\begin{figure}[ht!]
 \centering
 \includegraphics[scale=0.2,clip]{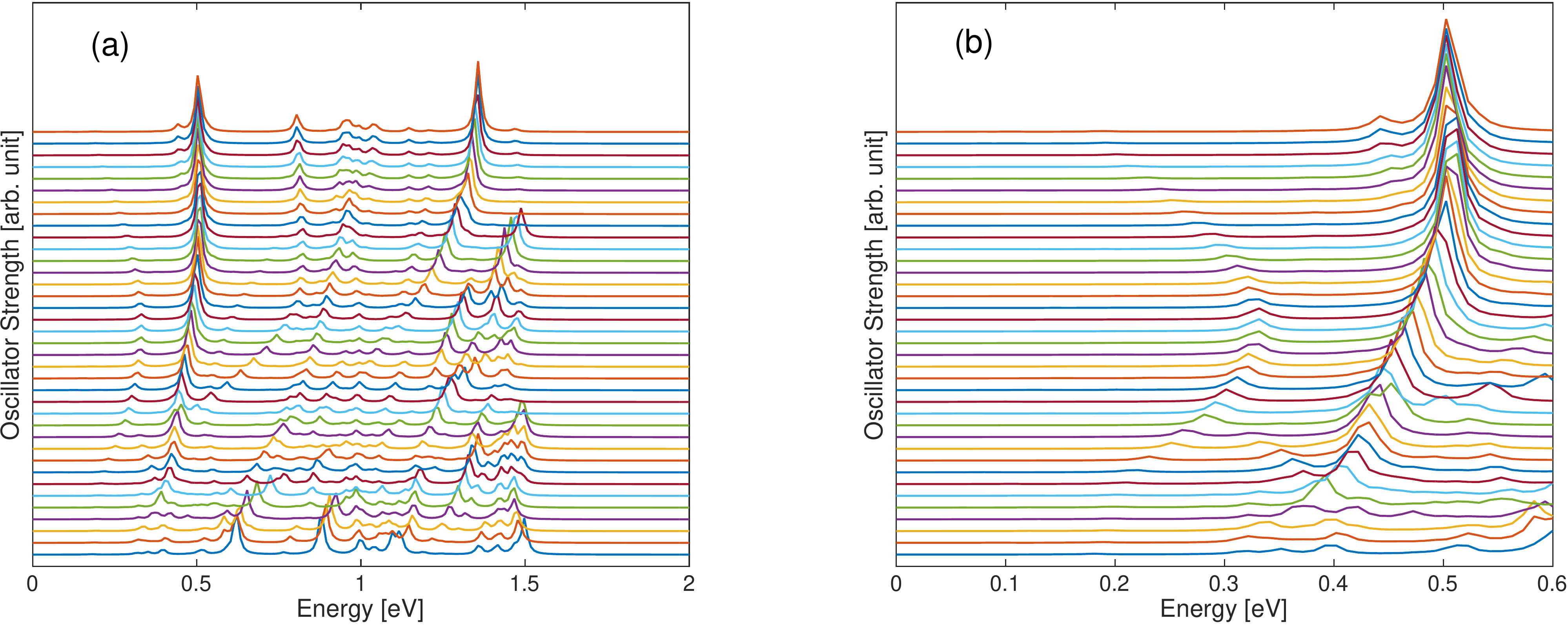}
 \includegraphics[scale=0.135,clip]{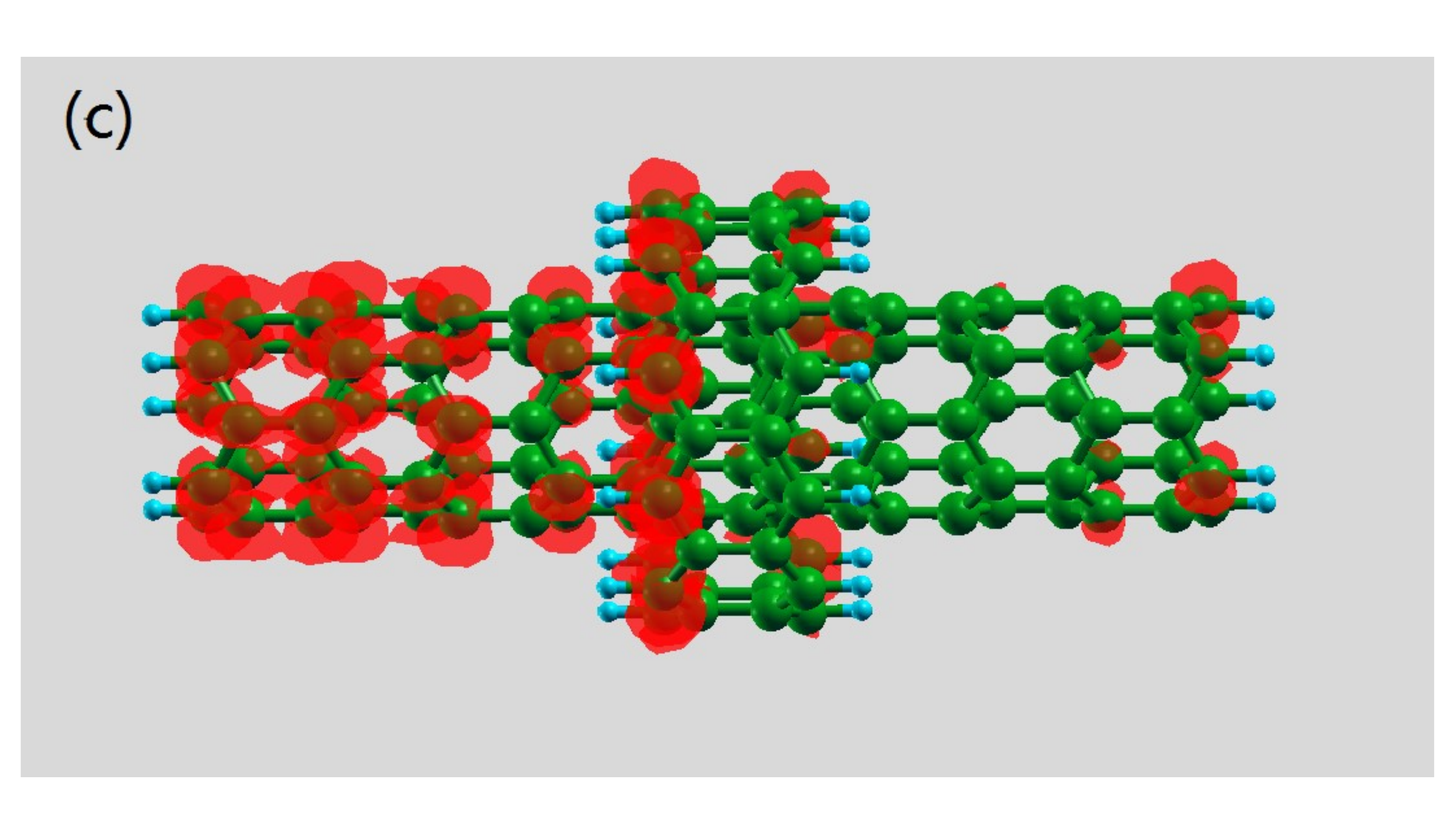}
 \includegraphics[scale=0.135,clip]{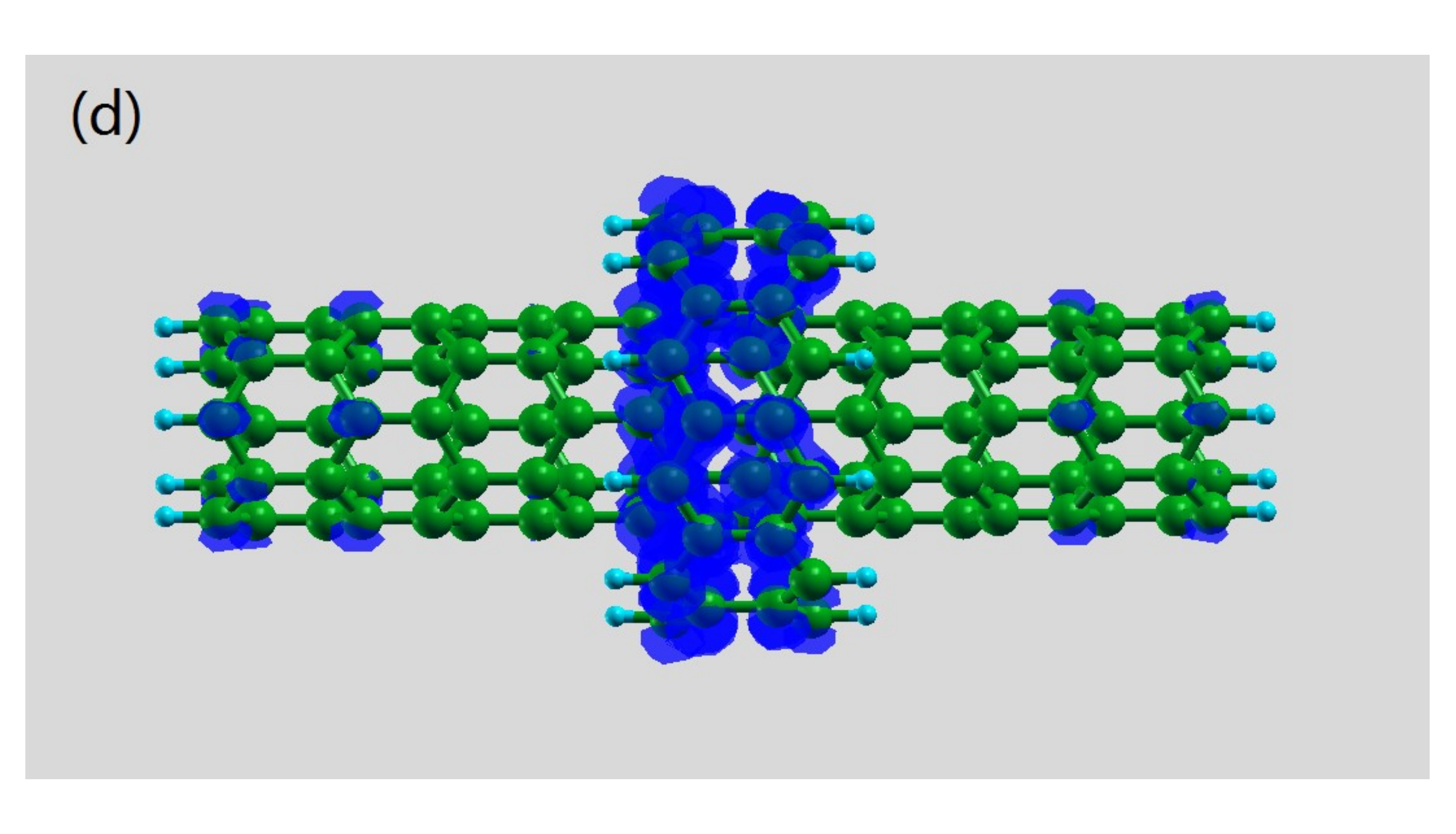}
 \caption{(Color online) (a)(b): the optical absorption spectrums of the nanomotor made from (5,0)/(10,0) DWCNT when the outer tube rotates around the inner tube counterclockwise. Here, the translation distance $R$ is fixed as $8.53$~\AA,~and the rotation angle varies from $0^{\circ}$ to $18^{\circ}$ with step length $0.5^{\circ}$ (shown from bottom to top). The spectrums are broaden by Lorentz lineshape with width $0.02$~eV. (c)(d): charge density distributions of the initial (red) and final (blue) state of the transition corresponding to the strong peak when $\Phi=2^{\circ}$ in (b), and the isovalue is taken as $5\times10^{-4}$.}
 \label{rotopt}
\end{figure}

Lastly, we study the evolution of the optical absorption spectrum of the nanomotor when the outer tube rotates around the inner tube counterclockwise. The nanomotor used here is made from (5,0)/(10,0) DWCNT, with one unit cell for the outer tube and five unit cells for the inner tube. The structure of this nanomotor has five-fold rotation symmetry around the axis, and the outer tube itself has ten-fold rotation symmetry. Therefore, the structure and optical absorption spectrum will repeat with the period $36^{\circ}$ when the outer tube continuously rotates around the inner tube. Besides, the configurations to be calculated can be reduced to the range $\Phi\in [0^{\circ},18^{\circ}]$, if we further consider the reflection operation of the nanomotor over one plane passing through the tube axis and $\Phi=0^{\circ}$. In Fig.~\ref{rotopt} (a) and (b), we show the calculated optical absorption spectrum of the nanomotor in different energy range when the rotation angle $\Phi$ is increased from $0^{\circ}$ to $18^{\circ}$ with step length $0.5^{\circ}$ and the translation distance $R$ is fixed as $8.53$~\AA. In contrast to the case of translational motion, we find that most peaks in the spectrum evolve smoothly and regularly during the rotational motion of the nanomotor(Fig.~\ref{rotopt}(a)), which implies less importance of quantum confinement effect here. Besides, some peaks appear in the spectrum and then shift during the rotational motion of the nanomotor. As shown in Fig.~\ref{rotopt}(b), a peak located above $0.3$~eV gradually appears and its intensity becomes stronger and stronger when $\Phi$ varies from $0^{\circ}$ to $18^{\circ}$, accompanying with the monotonic blueshift of its energy. The amplitude of the blueshift of the peak is as large as $200$~meV, which can also be easily detected experimentally.\cite{NatNano2013,NatPhys2014,PNAS2014} We also notice that the shift of the peak becomes slow when the rotation angle $\Phi$ approaches $18^{\circ}$, where the distances between the carbon atoms in the outer and inner tubes become relatively larger and their interaction becomes smaller. Besides, we plot the charge density distributions of the initial and final states of the corresponding transition when $\Phi=2^{\circ}$ in Fig.~\ref{rotopt} (c) and (d) respectively. We see that the initial state is distributed over one half of the nanomotor structure; while the final state is almost located at the touching region of the inner and outer tubes, which can be modulated by the rotation of the outer tube. Therefore, our calculation results here confirm that it is also possible to monitor the rotation motion of the nanomotor by detecting its optical absorption spectrum.

\section{Conclusions}
In conclusion, we have calculated the optical absorption spectrum of the nanomotor made from the double-wall carbon nanotube with the time-dependent density functional based tight binding method. The calculation results suggest that the spectrum depends sensitively on the configuration of the nanomotor when there exists strong enough interaction between the inner and outer tubes. When the outer tube moves along the inner tube, there exists one peak the energy of which shifts with an approximate period in the spectrum, while the other peaks don't shift periodically due to the quantum confinement effect of the ultrashort inner tube. When the outer tube rotates around the inner tube, most of the peaks in the spectrum shift smoothly and regularly. The shift amplitudes of the peaks in the spectrum are found to several hundred meV, which are easily detected by modern optical spectroscopy techniques.\cite{NatNano2013,NatPhys2014,PNAS2014}  Our studies here demonstrate the possibility to monitor the mechanical motion of the double-wall carbon nanotube based nanomotor from the evolution of its optical absorption spectrum, which may have potential applications in designing, fabricating, and utilizing nanoelectromechanical devices.

\section*{Acknowledgements}

We thank T.A. Niehaus for providing the TD-DFTB code and helpful discussions. This work was supported by NSFC Project No. 11604162, No. 61674083,
No. 51522201, and No. 11474006.

\bibliography{rsc} 
\end{document}